\documentclass[aps,pra,twocolumn,superscriptaddress,a4paper,10pt]{revtex4-2} 
\usepackage[T1]{fontenc}
\usepackage[utf8]{inputenc}
\usepackage{array}
\usepackage{multirow}
\usepackage{graphicx}
\usepackage{amsmath,amsfonts,amssymb,color}
\usepackage{amsthm}
\usepackage{mathtools}
\usepackage{leftidx}
\usepackage{xcolor}
\usepackage{dcolumn}
\usepackage{bm}
\usepackage{epstopdf}
\usepackage{epsfig}
\usepackage{environ}
\usepackage{pdfcomment}
\usepackage{float}
\usepackage{tabularray}
\usepackage{hyperref}
\usepackage{soul}
\usepackage{siunitx}

\usepackage{esint}
\usepackage{physics}
\usepackage[english]{babel}

\hypersetup{colorlinks=true,citecolor=blue,
	linkcolor=blue,urlcolor=blue,pdfstartview=FitH,
	bookmarksopen=true}
\usepackage{cleveref}

\begin{document}

	\title{Time-optimal force sensing with ultracold atoms}
	
\author{N. Ombredane}
\affiliation{Laboratoire Collisions Agrégats Réactivité, Université de Toulouse, CNRS, 31062 Toulouse, France}

\author{E. Flament}
\affiliation{Laboratoire Collisions Agrégats Réactivité, Université de Toulouse, CNRS, 31062 Toulouse, France}

\author{C. Babin}
\affiliation{Universit{\'e} Bourgogne Europe, CNRS, Laboratoire Interdisciplinaire Carnot de Bourgogne ICB UMR 6303, 21000 Dijon, France}

\author{D. Sugny}
\affiliation{Universit{\'e} Bourgogne Europe, CNRS, Laboratoire Interdisciplinaire Carnot de Bourgogne ICB UMR 6303, 21000 Dijon, France}

\author{D. Guéry-Odelin}
\affiliation{Laboratoire Collisions Agrégats Réactivité, Université de Toulouse, CNRS, 31062 Toulouse, France}

\author{B. Peaudecerf}\email[Corresponding author: ]{bruno.peaudecerf@cnrs.fr}
\affiliation{Laboratoire Collisions Agrégats Réactivité, Université de Toulouse, CNRS, 31062 Toulouse, France}
    
    \date{\today}

\begin{abstract}

We develop a time-optimal approach to force sensing using a Bose–Einstein condensate in a shaken optical lattice. Optimal control protocols are derived from a Fisher information framework and yield optimal dynamics that spontaneously organize in interferometer-like structures, where multiple interferences combine to maximize sensitivity. We analyze how measurement precision scales with control time and how the finite momentum dispersion of the condensate changes the optimal dynamics, observing an abrupt change of conformation from single- to double-folded interference structures for robust controls. The protocols are implemented experimentally for cold atoms subjected to inertial and magnetic forces, demonstrating high sensitivity and robustness. Our approach establishes a general route to time-optimal quantum sensing beyond standard interferometric architectures, applicable across all quantum platforms.
\end{abstract} 
   
\maketitle

\noindent\textit{Introduction.} 
Quantum sensing is one of the most advanced areas of quantum technologies~\cite{Acin_2018}, and has already resulted in practical applications~\cite{Degen_2017,Pezze_2018,Rembold_2020}. The goal of the field is to exploit a quantum system’s sensitivity to a physical quantity to measure it with unprecedented precision. Ultimate detection limits can be achieved in metrology measurements that use entanglement and rely on large amounts of data~\cite{Giovannetti_2004,Giovannetti_2011}. Another perspective, based on more practical processes, aims at measuring a quantity as precisely as possible with a limited number of resources, through a single or a few manipulations of quantum systems over a short period of time. In other words, the goal of this second approach, given sensitivity and experimental constraints, is to fully exploit the capabilities of the experimental apparatus to estimate the value of a physical quantity in a minimum time. Quantum optimal control (QOC), which is a set of tools for designing time-dependent controls to perform specific quantum operations in the best possible way while accounting for experimental limitations and uncertainties~\cite{Khaneja_2005,Amri_2019,Boscain_2021,Koch_2022,Ansel_2024}, provides a systematic mean of achieving this objective. 

\begin{figure}[ht!]
    \centering
    \includegraphics[width=0.8\columnwidth]{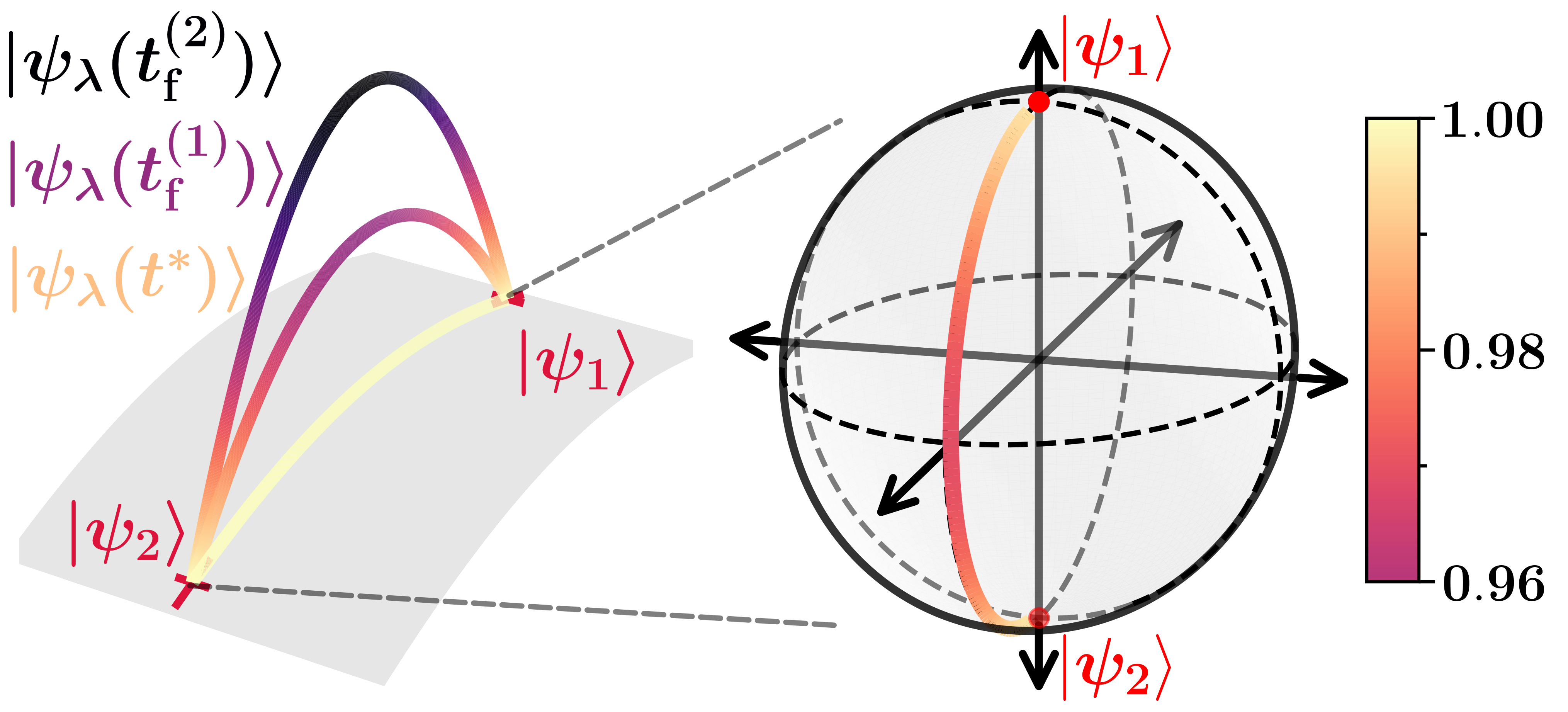}
    \caption{(Left) Schematic illustration of the convergence of the fingerprinting approach to a geodesic curve $\ket{\psi_\lambda(t^*)}$ between the target states $|\psi_1\rangle$ and $|\psi_2\rangle$. For a given parameter interval $\delta\lambda$, the time $t^*$ is the minimum control time required to reach the two states for two systems characterized by the values $\lambda_1$ and $\lambda_2$ chosen for fingerprinting. The times $t^{(1)}_{\rm f}$ and $t^{(2)}_{\rm f}$ verify $t^{(2)}_{\rm f}>t^{(1)}_{\rm f}>t^*$. (Right) Representation of an optimal control curve $\{\ket{\psi_\lambda(t^*)}\}$ on the Bloch sphere of the subspace generated by the states $|\psi_1\rangle$ and $|\psi_2\rangle$. The color of the curve denotes the fraction of the final state  contained in the subspace. The control protocol producing this curve was derived for force sensing with a BEC in a lattice (see main text), and approaches a geodesic curve, which corresponds to a great circle between the poles of the Bloch sphere.}
    \label{fig:Geodesic}
\end{figure}

 When estimating a continuous Hamiltonian parameter, the Cramer-Rao (CR) bound~\cite{Helstrom_1973}, which is defined in terms of the Classical Fisher Information (CFI), gives a lower bound to the sensitivity of a measurement protocol~\cite{Braunstein_1994,Latune_2013,Penasa_2016}. In a quantum measurement, the CFI depends on the Positive-Operator Valued Measure (POVM) describing the measurement process. The Quantum Fisher Information (QFI), which is the maximum CFI over all possible measurements, provides  a sensitivity bound for a given Hamiltonian evolution~\cite{Liu_2020}. A natural goal in metrology is then to find conditions that maximize the QFI, through a choice of control or of initial states~\cite{Liu_2022,Penasa_2016}. However, since the corresponding measurement may not be feasible in a given device, the QFI cannot generally be saturated experimentally. Therefore, a practical optimization should instead focus on the CFI. 

In this work, we carry out this general procedure with the aim of determining the optimal protocol to estimate the value of a constant force $f$ acting on a trapped quantum gas~\cite{Morsch_2006,Vanfrank_2016,Geiger_2020}
with a given level of precision in the shortest possible time. Our unique approach is to derive physical limits for the global estimation process, without assumptions on the dynamical structures, in contrast to previous approaches where only the elementary unitary operations of a pre-determined interferometer were optimized~\cite{Butts_2013,Weidner_2018,Rodzinka_2024,Saywell_2020,Saywell_2023,Ledesma_2025,Ledesma_2024}. 
We consider $\rm ^{87} Rb$ atoms in a shaken one-dimensional optical lattice, where the quantum state can be manipulated with the lattice position as the only control parameter~\cite{Dupont_2021}, and subjected to inertial or magnetic forces. 

 Rather than attempting to directly optimize the CFI, we employ an efficient and robust numerical method, the Fingerprinting~(Fp) approach ~\cite{Ma_2013,Ansel_2017}, which involves here maximizing the response difference of the system for two values of force separated by $\delta f$, the required sensitivity. The target states for Fp are chosen as momentum components of the BEC in a lattice, which can be directly measured. 
For the time-optimal protocol, we demonstrate that this approach is equivalent to CFI maximization for a specific measurement associated to the chosen target momenta. More specifically, we show that the set of final states parameterized by the force $f$ in the interval $\delta f$ forms a geodesic curve in the Hilbert space linking the two target states used for Fp. On this curve, the CFI is constant and equal to the QFI. This intrinsic geometric result illustrated in Fig.~\ref{fig:Geodesic} is general and applies to any parameter estimation in a quantum system.

The resulting optimal momentum dynamics are reminiscent of traditional interferometer structures, with splitting and recombination of wave amplitudes, which emerge naturally from the optimization process.
We then study the minimum time $t^*$, showing that it scales as $\delta f^{-1/3}$, in line with the scaling of accelerated sequential interferometers~\cite{Storey_1994,Mcdonald_2013,Mcdonald_2014}.
However, our results stand in contrast with sequential protocols---which require mitigating the impact of imperfect elementary pulses~\cite{Beguin_2023,Rodzinka_2024}---in that our optimization approach globally ensures that the full dynamics only results in populating two output modes. We also emphasize that in our approach, the lattice potential is always present, enabling continuous control of the BEC, in a regime far from the two-level approximation~\cite{Saywell_2020}.
 
In an experimental context, a crucial point is to account for the finite size of the system, which results in momentum dispersion. The control then needs to robustly provide the same outcome over a range of momenta $\Delta q$. We find that the robust dynamics undergo a sharp transition between a single-fold momentum interferometric structure for low dispersions---typically given by $\Delta q \lesssim  \delta f t^*$---and for larger momentum widths, a doubly-folded structure corresponding to a spatial recombination of wave packets which ensures robustness. We finally apply such robust protocols experimentally to the measurement of inertial forces and the mapping of local magnetic field gradients.

\begin{figure}[ht!]
    \centering
    \includegraphics[width=\columnwidth]{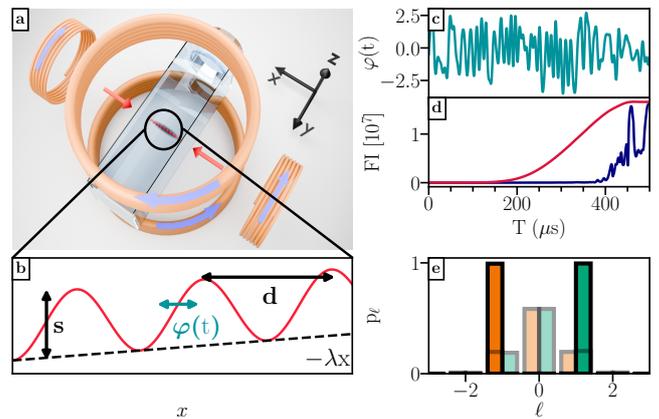}
    \caption{\textbf{(a)}. Experimental setup: the BEC is produced in a glass cell and trapped in an optical lattice potential produced by counter-propagating beams (red arrows) and subjected to the combined magnetic field of quadrupole coils along the $z-$axis, and horizontal coils along the lattice $x-$axis that allow to displace the zero of the field. \textbf{(b)} The lattice potential is characterized by its spacing $d$ and depth $s$. The control parameter $\varphi(t)$ translates the lattice along $x$, and an external constant force $\lambda$ is applied. \textbf{(c-e)}  Typical numerical optimization results: \textbf{(c)} control $\varphi(t)$, \textbf{(d)} evolution of the quantum (red) and classical (blue) Fisher informations, and \textbf{(e)} final probability distribution over the states $\ket{\ell}$ for $\lambda=\lambda_1$ (orange) and $\lambda_2$ (green). The corresponding final distributions obtained in the absence of control are superimposed in faded colors. Parameters: $\lambda_1 = 0, \lambda_2 = 7.5\cdot 10^{-4}$.}
    \label{fig:setup}
\end{figure}

\noindent\textit{Principle.}
Consider a quantum system described by a Hamiltonian $\hat{H}_\lambda$ that depends on a parameter $\lambda$ to be estimated, with a value in the vicinity of $\lambda_0$. The system undergoes dynamics driven by a time-dependent control $u(t)$ for a duration $t_{\rm f}$, with a final state $|\psi_\lambda(t_{\rm f})\rangle$, solution of the Schr\"odinger equation (in units where $\hbar=1$)
$
i|\dot{\psi}_\lambda(t)\rangle = \hat{H}_\lambda(u)|\psi_\lambda(t)\rangle,
$
starting from an initial state independent of the exact value of the unknown parameter, i.e. $|\psi_\lambda(0)\rangle=|\psi_{\lambda_0}(0)\rangle$. 
For a single measurement, the mean square error $\Delta\lambda$ on the estimation of $\lambda$ from quantum measurements on the system is bounded below  by the inverse of the QFI $\mathcal{F}^{(Q)}$ as $\Delta\lambda\geq \tfrac{1}{\sqrt{\mathcal{F}^{(Q)}}}$. Expanding the state in Taylor series,   $|\psi_\lambda(t_{\rm f})\rangle = |\psi_{\lambda_0}(t_{\rm f})\rangle +|\partial_\lambda\psi\rangle (\lambda-\lambda_0) +O((\lambda-\lambda_0) ^2)$, with $|\partial_\lambda\psi\rangle$ the state derivative, the QFI can be expressed 
as:
$$
\mathcal{F}^{(Q)}_{\lambda_0}=4\left(\langle \partial_\lambda \psi|\partial_\lambda\psi\rangle-
|\langle\psi_{\lambda_0}|\partial_\lambda\psi\rangle |^2\right),
$$
i.e. four times the Fubini-Study (FS) metric. This provides a direct geometric interpretation of the QFI, as the distance separating two final states under an infinitesimal variation of $\lambda$.

To minimize $\Delta\lambda$, we use Fp, which forces the outcome of the dynamics onto two measurable states denoted by $|\psi_1\rangle$ and $|\psi_2\rangle$, with $\langle\psi_1|\psi_2\rangle=0$, associated with values $\lambda_{1,2}=\lambda_0\mp\delta\lambda/2$ for the parameter $\lambda$. The Fp is defined from the figure of merit $\mathsf{F}_{\lambda_1,\lambda_2}$ given by
\begin{equation}\label{eqF}
\mathsf{F}_{\lambda_1,\lambda_2}(t_{\rm f})=\frac{1}{2}\left(|\langle\psi_{\lambda_1}(t_{\rm f})|\psi_1\rangle|^2+|\langle\psi_{\lambda_2}(t_{\rm f})|\psi_2\rangle|^2\right),
\end{equation}
 maximized by the simultaneous optimal control of the dynamics with parameter values $\lambda_1$ and $\lambda_2$. The maximum value $\mathsf{F}_{\lambda_1,\lambda_2}=1$ is achieved for large enough control times $t_{\rm f}$, whose minimum $t^*$ corresponds to the quantum speed limit of the protocol~\cite{Deffner_2017,Ness_2021}. Conversely, for a fixed time $t_{\rm f}$ there is a minimal parameter difference $\delta\lambda^*$ for which $\mathsf{F}_{\lambda_1,\lambda_2}=1$ can be satisfied. 

Given a control time $t_{\rm f}$ and fixed values $\lambda_1$ and $\lambda_2$, we  consider the trajectories in Hilbert space defined by the set of states $\{|\psi_\lambda(t_{\rm f})\rangle\}$ for $\lambda_1\leq \lambda\leq \lambda_2$, achieved by a control $u(t)$ for which $\mathsf{F}_{\lambda_1,\lambda_2}(t_{\rm f})=1$. Their length $\ell(t_{\rm f})$ can be written generally as
\begin{equation}\label{eqlength}
\ell(t_{\rm f})=\int_{\lambda_1}^{\lambda_2}ds=\int_{\lambda_1}^{\lambda_2}\frac{\sqrt{\mathcal{F}^{(Q)}_\lambda(t_{\rm f})}}{2}d\lambda,
\end{equation}
where $ds$ is an infinitesimal arc length from the FS metric. By construction, $\ell(t_{\rm f})\geq \tfrac{\pi}{2}$ where $\tfrac{\pi}{2}$ is the length of the geodesic between the orthogonal states $|\psi_1\rangle$ and $|\psi_2\rangle$. In the fixed control time $t_{\rm f}$ and for a finite-dimensional Hilbert space, the QFI is upper bounded~\cite{Pang_2017} (See Appendix~\ref{secCoincidences} for details). We denote $\mathcal{F}_{\rm max, \mathsf{F}=1}^{(Q)}$ the bound for the trajectories considered, to get
\begin{equation}
    \frac{\pi}{2}\leq\ell(t_{\rm f})\leq\frac{\sqrt{\mathcal{F}_{\rm max, \mathsf{F}=1}^{(Q)}(t_{\rm f})}}{2}\delta\lambda.
\end{equation}
The smallest achievable value of $\delta\lambda$ must therefore correspond to a control for which $\ell(t_{\rm f})$ saturates both bounds, and the states $\{|\psi_\lambda(t_{\rm f})\rangle\}$ thus define a geodesic in Hilbert space between states $\ket{\psi_1}$ and $\ket{\psi_2}$ (see Fig.~\ref{fig:Geodesic} for a schematic representation). We also deduce that the QFI is a constant function of $\lambda$ along the geodesic, with $\mathcal{F}_{\rm max, \mathsf{F}=1}^{(Q)}(t_{\rm f})=\tfrac{\pi^2}{\delta\lambda^{*2}}$. Furthermore, on this curve the QFI can be shown to coincide with the CFI from a projective measurement that includes projectors on $\ket{\psi_{1,2}}$ (Appendix~\ref{secCoincidences}).
This intrinsic geometric property is valid for any parameter-dependent quantum dynamics. It connects a practical estimation method based on Fp to a fundamental sensing limit given by the Fisher information.

\noindent\textit{Characterization of the optimal force sensing protocols.}
We now focus on the goal of measuring a constant force, represented by a potential $\hat{V}_\lambda=-\lambda \hat{x}$, from the dynamics of a quantum particle in a periodic potential $\hat{V}_{\rm L}=-(s/2)\cos(\hat{x}+\varphi(t))$, with $\hat{x}$ and $s$ the dimensionless position coordinate and depth of the lattice, respectively. This can be realized experimentally by subjecting cold atoms trapped in a lattice potential to a magnetic field gradient, as sketched in Fig.~\ref{fig:setup}\textbf{a}.  The phase $\varphi(t)$ acts as the control parameter $u(t)$. In the interaction representation with respect to $\hat{V}_\lambda$, the dynamics are governed by the Schrödinger equation
\begin{equation}\label{eqSchro}
i\ket{\dot{\psi}} = \left[ (\hat{p}-\lambda t)^2-\frac{s}{2}\cos(\hat{x}+\varphi(t))\right]\ket{\psi}.
\end{equation}

\begin{figure}[ht!]
    \centering
        \includegraphics[width=\linewidth]{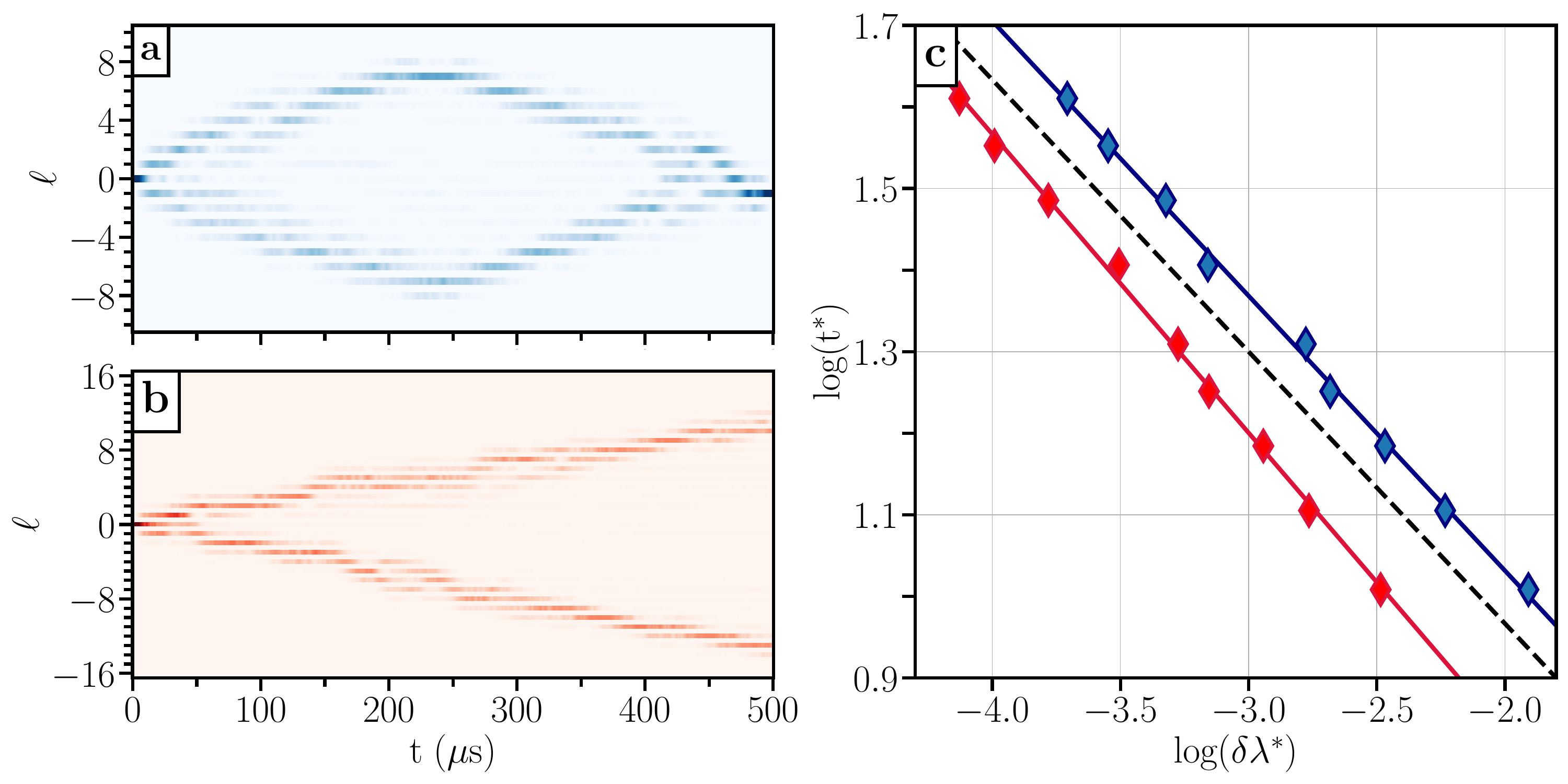}
    \caption{{\bf (a)} Emerging interferometer pattern in the optimal time evolution of the momentum population in an infinite lattice with $q=0$, for $\lambda_1=0$ and $\lambda_2 = 7\cdot10^{-4}$. The dynamics are shown for $\lambda=\lambda_1$. {\bf (b)} Optimal trajectory for a direct optimization of the QFI (see text).  
    {\bf (c)} Evolution of the minimum time $t^*$ as a function of $\delta \lambda$ for Fp (blue diamonds) and a direct maximization of the QFI (red diamonds). Numerical uncertainties are typically smaller than the symbol size. Linear regressions are shown in blue and red lines, yielding slopes of  $-0.34 \pm 0.01$ and $-0.37 \pm 0.01$ respectively. The scaling $t^*\propto(\delta\lambda^*)^{-1/3}$ is shown by the dashed black line.}
    \label{fig:theorie_Tstar}
\end{figure}

The resulting dynamics are invariant by translation of the lattice period ($2\pi$). The quasi-momentum $q\in (-0.5,0.5]$ is therefore conserved and we can solve for the dynamics separately for each value of $q$~\cite{Dupont_2021,Ansel_2024}. In our experiments, the initial state is the ground state $\ket{\psi_0}$ of the lattice, for which $q=0$, and the corresponding subspace of Hilbert space is spanned by the discrete momentum basis $\ket{\ell}$
with $\braket{ x}{\ell} =\tfrac{1}{\sqrt{2\pi}}e^{i\ell x}$, $\ell\in\mathbb{Z}$. 
In this ideal case, we implement the Fp strategy with the target states $\ket{\psi_1} = \ket{-1}$ and $\ket{\psi_2} = \ket{1}$. A gradient-based algorithm is used to determine a control protocol that maximizes the fidelity of Eq.~\eqref{eqF} in minimum time $t^*$~\cite{Dupont_2021,Ansel_2024}. A typical control is depicted in Fig.~\ref{fig:setup}\textbf{c}.  We obtain a coincidence of QFI and the CFI defined from the measurement of populations in the states $\{\ket{\ell}\}$ at the end of the control (Fig.~\ref{fig:setup}\textbf{d}). The final orthogonal states obtained for $\lambda_1$ and $\lambda_2$ have to be contrasted with those generated by the control $\varphi(t)=0$, which are almost indistinguishable from a sole probability measurement (Fig.~\ref{fig:setup}\textbf{e}). The resulting evolution of the final states $\ket{\psi_\lambda(t_{\rm f})}$ is very close to a geodesic between $\ket{\psi_1}$ and $\ket{\psi_2}$ (see Fig.~\ref{fig:Geodesic}). 

\begin{figure*}
    \centering
    \includegraphics[width=\linewidth]{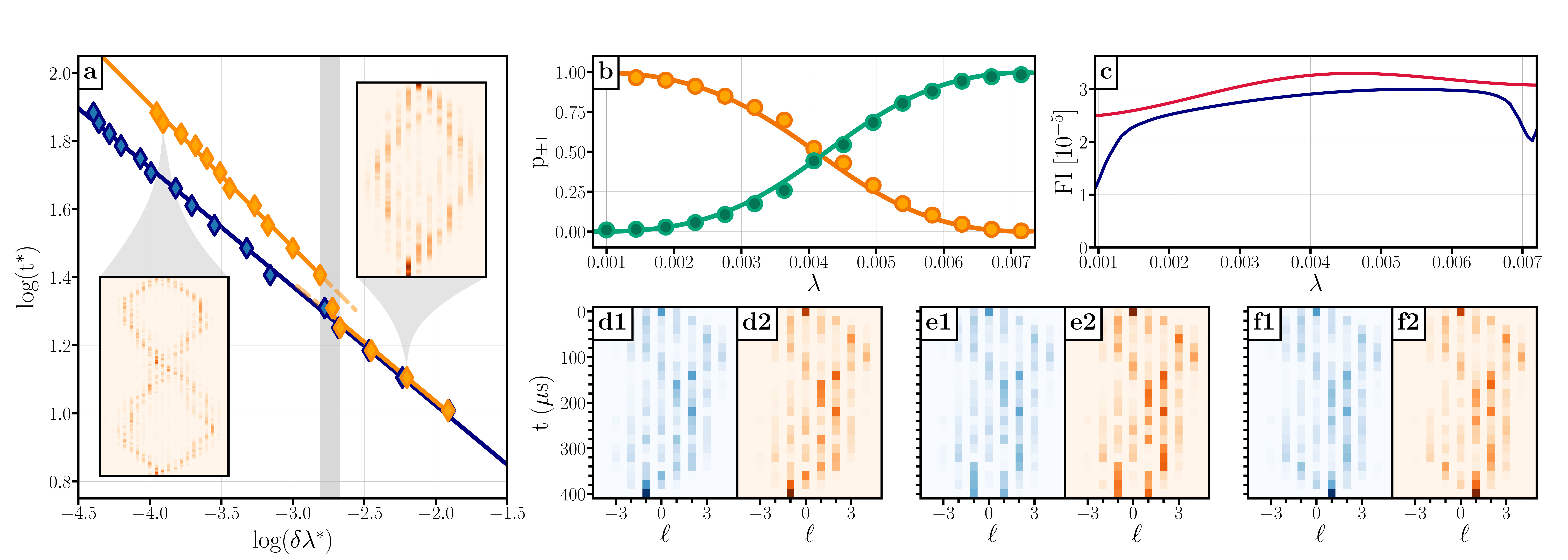}
    \caption{\textbf{(a)} Evolution of the minimum time $t^*$ as a function of $\delta \lambda$ for the Fp approach (blue diamonds) at $q=0$, with $\lambda_1=0$ and a quasi-momentum robustness constraint with $\Delta q=0.002$ (yellow diamonds). With the robustness constraint, a sharp change of behaviour occurs (grey stripe). Insets show typical momentum distribution evolutions in the two regimes of the robust case. Linear regressions are shown in blue and yellow lines, yielding slopes of  $-0.349(4)$ (blue) and $-0.35(3)$ and $-0.42(1)$ (yellow, for low and high values of $\delta\lambda^*$, respectively). \textbf{(b-d)} Experimental demonstration of a robust Fp control with $\lambda_1=10^{-3}$, $\lambda_2=7.15\cdot10^{-3}$, $t^*=400$ \unit{\micro\second}, $\ket{\psi_1}=\ket{-1}$, $\ket{\psi_2}=\ket{1}$ and $\Delta q=0.06$. \textbf{(b)} Measured populations $p_{\pm1}$ in states $\ket{\pm1}$ as the inertial force $\lambda$ is varied. Measured standard errors on probability are smaller than marker size. Full lines show the numerical prediction, including a correction for the impact of scattering halos during imaging (see Appendix~\ref{app:exp}). \textbf{(c)} Numerically computed Quantum (upper red curve) and Classical (lower blue curve) Fisher information. \textbf{(d-e)} Experimentally measured (blue, panel 1) and numerical (red, panel 2) momentum distribution evolutions for $\lambda=\lambda_1$, $\lambda_0$ and $\lambda_2$ respectively.}
    \label{fig:acc}
\end{figure*}

The optimization protocol does not provide any a priori knowledge on the way
the minimum time $t^*$ depends on the interval $\delta \lambda $, nor on the time-dynamics of the quantum state for the different values of $\lambda$. We investigate these aspects numerically. A typical evolution is depicted in Fig.~\ref{fig:theorie_Tstar}\textbf{a}: the state of the system is shown to split into two opposite accelerated branches, that merge again at the end of the dynamics. 
This is reminiscent of an interferometer structure, which we emphasize emerges here from the sole unsupervised Fp optimization process. In Fig.~\ref{fig:theorie_Tstar}\textbf{c}, we investigate how $t^*$ varies with $\delta \lambda$ and find good agreement with the scaling $t^* \propto \delta\lambda^{-1/3}$. 
This scaling is the same as that obtained from a direct optimization of the QFI, and can be understood from the accelerated-decelerated interferometric structure of the dynamics, which constitutes a \emph{ladder-climbing} process along symmetric momenta superposition states that maximize the QFI at a given energy (see Appendix~\ref{scalingQFI}). A sole optimization of the QFI $\mathcal{F}_{\lambda_0}^{(Q)}$, results in the dynamics of Fig.~\ref{fig:theorie_Tstar}\textbf{b}, where the system evolves along increasingly large superpositions leading to a similar scaling and smaller $t^*$ values (Fig.~\ref{fig:theorie_Tstar}\textbf{c}). However, in that circumstance, a coincidence between QFI and CFI is not guaranteed since an appropriate measurement may not be available, as opposed to the Fp approach.

In practice,  
the momentum dispersion $\Delta q$ of the BEC cannot be neglected in experiments and numerical simulations.
Keeping the same initial and target states within quasi-momentum classes, we design robust control protocols with respect to the quasi-momentum for different values of $\Delta q$. The robust control is designed through the simultaneous optimization of an ensemble of systems characterized by different values of $q$, and governed by the Hamiltonian
\begin{align}
\hat{H}_q&=\sum_\ell(\ell+q-\lambda t)^2\ket{\ell}\bra{\ell} \label{eqSchroQ}\\
&-\frac{s}{2}\left(e^{i\varphi(t)}\ket{\ell}\bra{\ell+1}+e^{-i\varphi(t)}\ket{\ell+1}\bra{\ell}\right).\nonumber
\end{align}

Numerically, we consider a regular distribution of 5 values of $q$ in the range $[-\Delta q/2,\Delta q/2]$. A comparison of the computed $t^*$ when introducing robustness over a range $\Delta q=2\cdot 10^{-3}$ is shown on Fig.~\ref{fig:acc}\textbf{a}. While the scaling in the non-robust case obeys $t^*\propto(\delta\lambda^*)^{-1/3}$ over the whole range of $\delta\lambda$, a sudden jump in the $t^*$ curve accompanied by a slight change in slope ($t^*\propto(\delta\lambda^*)^{-0.4}$) is obtained when introducing robustness. This shift of the scaling time is accompanied by a topological change in the interferometric structure, resulting in a doubly-folded pattern, which emerges naturally from the optimization process to ensure robustness with respect to the quasi-momentum in a process akin to spin-echo. This pattern is reminiscent of the shape of usual accelerated interferometers~\cite{Rodzinka_2024}. The tipping point between behaviors occurs as the product $\delta\lambda\, t^*$ gets small compared to $\Delta q$, which can be deduced from a qualitative analysis of Eq.~\eqref{eqSchroQ}. 

\noindent\textit{Experimental demonstration.}~We implement the robust optimal protocols on a $^{87}\rm Rb$ BEC containing about $\simeq 5\cdot 10^5$ atoms, loaded in the ground state of a one-dimensional optical lattice potential
$
\hat{W}_{\rm L}=-\frac{s E_{\rm L}}{2}\cos(k_{\rm L} \hat{X}+\varphi(T)),
$
where $k_{\rm L} = 2\pi/d$ and $E_{\rm L}=h^2/(2md^2)$ (with $m$ the mass of an atom and $h$ the Planck constant) are a characteristic wave vector and energy associated with the lattice, respectively (see Appendix~\ref{app:exp}). We recover the potential $\hat{V}_{\rm L}$ discussed above as $\hat{V}_{\rm L}=\hat{W}_{\rm L}/E_{\rm L}$, with dimensionless quantities $x=k_{\rm L}X$, $p=P/(\hbar k_{\rm L})$, $t=E_{\rm L}T/\hbar$. The control parameter $\varphi(T)$ is set by the relative phase between two acousto-optic modulators~(Appendix~\ref{app:exp}). 
We experimentally measure the momentum distribution of the BEC by absorption imaging following a $35\,\rm ms$ time-of-flight. The distribution consists of regularly spaced peaks corresponding to the momentum components $p\simeq\ell$.

As a first demonstration, we subject the BEC to a  well-controlled inertial force in the lattice, separating the phase modulation in two terms:
$
\varphi(T)=\varphi_{\rm Fp}(T)+\varphi_f(T).
$
The phase $\varphi_f(T)=\pi fT^2/(md)$ corresponds to a constant acceleration of the lattice. This emulates an inertial force $f$  in the reference frame of the lattice, with $\lambda=f/(k_{\rm L}E_{\rm L})$. The phase $\varphi_{\rm Fp}(T)$ implements the optimal control from Fp ensuring robustness over a quasimomentum width  $\Delta q= 0.06$ at the depth $s= 5$. Furthermore, we take practical limitations in hardware bandwidth into account by a frequency filtering of the control (Appendix~\ref{app:exp}). Despite these constraints, we get an optimal control that experimentally realizes the expected geodesic between $\ket{\psi_1}$ and $\ket{\psi_2}$.  
In Fig.~\ref{fig:acc}\textbf{b}, we show the evolution of the measured populations $P_{1,2}(\lambda)$ in states $\ket{\psi_{1,2}}$ as the value of the inertial force $\lambda$ is varied. As expected, the evolution of the states $\ket{\psi_\lambda(t_{\rm f})}$  is close to a geodesic curve, leading to a sinusoidal evolution of the probabilities. The corresponding QFI and CFI are shown in Fig.~\ref{fig:acc}\textbf{c}. As expected, the QFI is near constant (within 8\%) over the interval $\delta\lambda$, and the CFI is near coincident. Residual imperfections are due to practical convergence thresholds of our algorithm (see Appendix~\ref{app:numerics}). The interval $\delta\lambda=6\cdot 10^{-3}$ probed in this experimental example corresponds to an acceleration of $27\%$ of $g$, and a full transfer in this interval is achieved in $t^*=400$ \unit{\micro\second}. To compare the data to the numerical results, we  account for the impact of interactions between atoms of the BEC during time-of-flight (Appendix~\ref{app:exp}). The resulting corrected curves (Fig.~\ref{fig:acc}\textbf{b}) show good agreement with the measured data, which validates our approach experimentally. 
Figure~\ref{fig:acc}\textbf{d-f} depicts the comparison of the expected and measured time-evolution of momentum distributions for values $\lambda_1, \lambda_0$ and $\lambda_2$, showing excellent match.
We can evaluate the precision $\Delta\lambda$ achieved with our protocol by estimating $\lambda$ from our measured data points. Although a reduction of $\Delta\lambda$ is observed when increasing the final CFI, residual experimental fluctuations contribute significantly to the measured uncertainties in this work.

\begin{figure}
    \centering
    \includegraphics[width=\columnwidth]{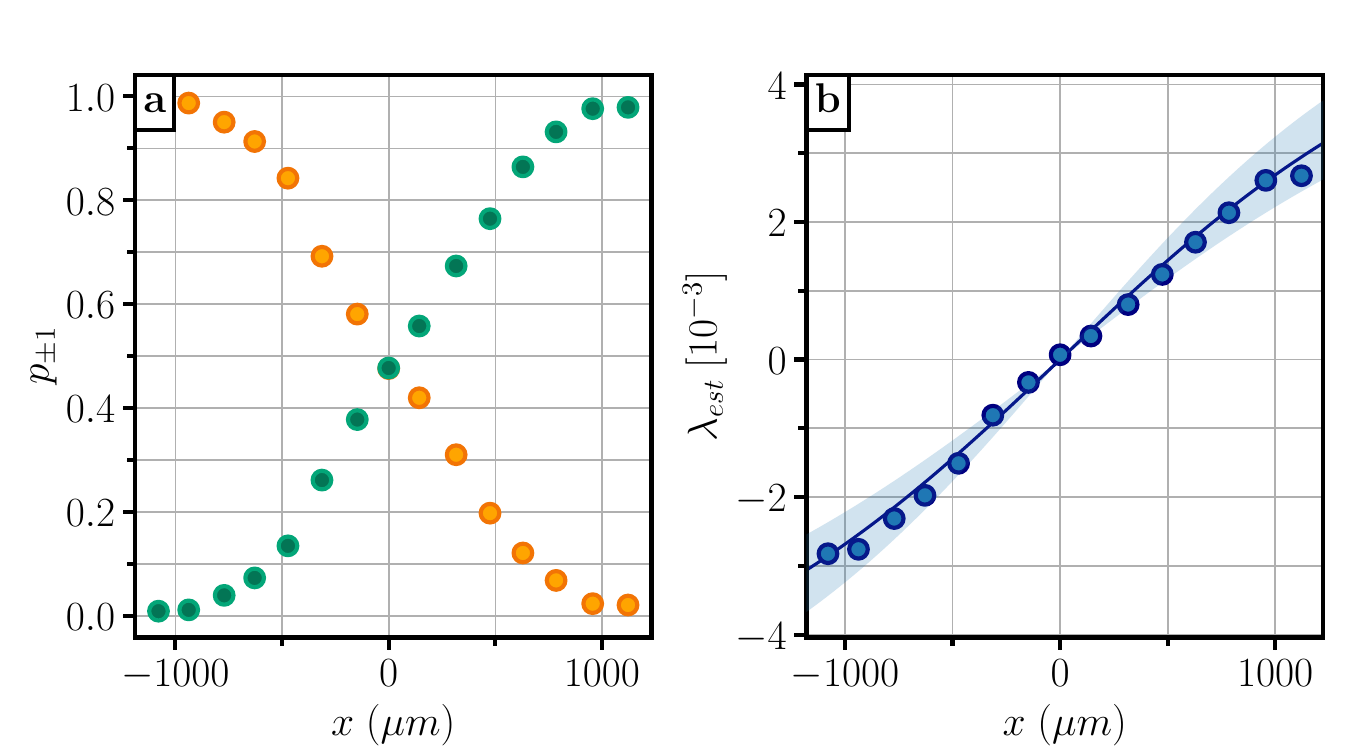}
    \caption{\textbf{(a)} Measured populations $p_{\pm 1}$ in states $\ket{\pm 1}$ after the Fp protocol, as the initial position of the condensate in the quadrupole magnetic field is varied. Measured standard errors on probability and position are smaller than marker size. \textbf{(b)} Deduced local estimate of the dimensionless force $\lambda$ (blue disks). The curve shows the expected behavior from an offset quadrupole field, with the light blue area denoting the uncertainties in fitting parameters.}
    \label{fig:force}
\end{figure}

Having demonstrated and benchmarked the Fp protocol in an experimental context, we finally apply it to characterize a local force gradient in the BEC setup. The magnetic quadrupole trap that helps produce the BEC (Appendix~\ref{app:exp}) also provides a vertical force compensating gravity during lattice experiments. As a consequence, a shallow trapping potential is also present along the $x-$axis. In the previous experiments relying on inertial forces, the total displacement of the  BEC over the Fp procedure was smaller than $d$, leading to negligible contributions of the magnetic potential to the measured force. To chart the spatial variations of the magnetic gradient, we start from the BEC, and displace it adiabatically over a millimetric distance, by moving the minimum of the magnetic potential via the progressive addition of a constant magnetic field along the $x-$axis (see Fig.~\ref{fig:setup}). The BEC is then adiabatically loaded in the ground state of the lattice, before the bias field is suddenly turned off. The displaced BEC is thus subjected to the local gradient from the non-biased magnetic trap. Following a holding time of  $200\,\unit{\micro\second}$ which allows the field to settle, we apply a control phase modulation $\varphi_{\rm Fp}(T)$ to measure the local force. The fingerprinting is optimized for states $\ket{\pm 1}$, and for $\lambda_1=-\lambda_2= 0.00308$. These measurements are shown in Fig.~\ref{fig:force}. The resulting force is mostly linear, as expected from the harmonic approximation at the bottom of the magnetic trap. We fit the data with the expected shape of the gradient from a quadrupole trap $\lambda(x)=\Lambda\frac{x}{\sqrt{x^2+4z_0^2}}$,
where $\Lambda$ is the asymptotic value of the force away from the trap center, and $z_0$ the vertical shift between the zero of the quadrupole field and the BEC. The fit yields values $\Lambda=6.4(6)\cdot 10^{-3}$ (corresponding to a gradient of 8.8(8) G/cm) and $z_0=1100\pm 100$ \unit{\micro\meter}, in reasonable agreement with independent measurements.

\noindent\textit{Conclusion.}~We have investigated the global optimization of controlled dynamics for parameter sensing in minimum time. Focusing on a fingerprinting strategy, we found that the global optimal solution is a geodesic in Hilbert space between the final states. We have applied this control strategy to force sensing with cold atoms in an optical lattice, where the optimal dynamics show the emergence of interferometer-like structures, which we understand as resulting from the generic form of states that optimize the QFI. We implemented our protocols experimentally, demonstrating the measurement of an inertial force, and the characterization of a local magnetic force gradient.

The interpretation of the fingerprinting strategy in terms of Fisher information combined with QOC provides a systematic and flexible sensing approach that can be applied to other quantum platforms. The optimization successfully tailors multiple interferences created by dynamics to ensure a final result that is exactly split between two output states. These optimal protocols offer a new perspective for fast inertial sensors based on atom interferometry, which could be extended in higher dimension to the sensing of force and rotation vectors. In that context, the impact of atom interactions on the optimal protocols remains to be investigated~\cite{Jager_2014,Dionis_2025}.

\paragraph{Acknowledgements.} This work was supported by the Institut Universitaire de France, the ANR project QuCoBEC (ANR-22-CE47-0008-02), and Thales Alenia Space. D. Sugny acknowledges the support of the CNRS projects QUSPIDE and CONV, and by the EUR EIPHI project “SQC.”
\paragraph{Data availability.} The data that support the findings of this Letter are available upon request. 

\clearpage
\bibliography{references2.bib}

\newpage

\appendix

\section{Experimental Methods}
\label{app:exp}

\noindent\textit{Lattice dynamics.} In the experiment, the $\rm ^{87}Rb$ atoms are subjected both to the lattice potential, and to the combined potential of a dipole trap beam along the $x$-axis and a magnetic quadrupole trap which form the hybrid trap where the BEC is produced. The hybrid trap potential is kept on during Fp experiments, and provides an additional weak harmonic trapping with frequencies $(\omega_x,\omega_y,\omega_z)=2\pi\times({ 10.3,63,78})$ Hz. Over the typical timescale of experiments ($<1\,\rm ms$), the impact of this additional confinement can be neglected.

The atoms of the BEC are polarized in the $\ket{F=1,m_F=-1}$ hyperfine sublevel of the ground state, where they interact through contact interactions characterized by a scattering length $a=104\,a_0$ with $a_0$ the Bohr radius. However due to the weak radial confinement the gas density remains low, and the system dynamics can accurately be described by the Schr\"odinger equation over the timescale of experiments. Our measurements however need to account for the impact of these interactions during the time-of-flight (see below).

\noindent\textit{Modulation constraints.} The optical lattice is formed from the interference of two beams that are split from a common laser source. Before the splitting, an acousto-optic modulator (AOM) allows to modulate the amplitude $s$ of the lattice with a bandwidth of $\simeq3\,$MHz. In this work the depth $s$ is kept constant and is independently calibrated before each experiment~\cite{Cabrera_2018}. After the main beam is split, each arm of the lattice goes through a second AOM, and phase modulation is applied on one arm. The bandwidth for phase modulation is in practice limited to frequencies below $1\,\rm MHz$, which is accounted for by a frequency filtering of the control: the control is written as a Fourier sum over the time $t_{\rm f}$ with a frequency cutoff set to $f_{\rm max}=125\,\rm kHz$ in the experiments presented here~\cite{Flament_2025}. The GRAPE algorithm then optimizes a discretized phase ramp, with constant steps of duration $0.5$ \unit{\micro\second}.

\noindent\textit{Scattering halos in time-of-flight measurements.} While interactions can be neglected over the timescale of the Fp dynamics, during the time-of-flight that precedes absorption measurement, atoms belonging to different components  of momentum $\ell\hbar k_{\rm L}$ can collide, leading to spherical collision halos~\cite{Chatelain_2020}. This modifies the final probability distribution over the measured peaks. In Fig.~\ref{fig:acc}b, the expected numerical probabilities $P_{1,2}(\lambda)$ are corrected to account for this effect.

\section{Numerical computation of minimum time controls}
\label{app:numerics}

The determination of the optimal-time scalings of Figs.~\ref{fig:theorie_Tstar} and~\ref{fig:acc} is performed numerically. For a given control time $t_{\rm f}$, a GRAPE algorithm is used to maximize the fidelity $\mathsf{F}$ (Eq.~\eqref{eqF}), for increasing values of the parameter interval $\delta\lambda$~\cite{Dupont_2021}. The optimization is set to run for a maximum of 2000 iterations or until the fidelity crosses the threshold of $99.5\%$. We found that better convergence was obtained by iteratively re-injecting the solution as an initial guess between increasing values of $\delta\lambda$.

This approach yields a monotonically increasing curve $\mathsf{F}(\delta\lambda)$ for a fixed control time $t_{\rm f}$, which typically rises quickly before reaching a plateau. Fitting this curve by the combination of a half inverted parabola and plateau allows to pinpoint $\delta\lambda^*$, the minimal parameter interval that can be resolved properly in the control time $t_{\rm f}$.

\section{Coincidence between QFI and CFI}\label{secCoincidences}

We establish here in a general setting the conditions for which the coincidence CFI=QFI may occur. We then deduce that the coincidence is expected at all times along the optimal curves in Hilbert space, which are geodesics.

We consider the evolution of a pure state $\ket{\psi_\lambda(t)}$ from state $\ket{\psi_0}$. As mentioned in the main text, the QFI with respect to $\lambda$ can be expressed as
$$
\mathcal{F}_\lambda^{(Q)}=4\left[\braket{\partial_\lambda\psi_\lambda(t)}{\partial_\lambda\psi_\lambda(t)}-|\braket{\psi_\lambda(t)}{\partial_\lambda\psi_\lambda(t)}|^2\right],
$$
where the state $\ket{\partial_\lambda\psi_\lambda(t)}$ is solution of
$$
i\frac{d}{dt}\ket{\partial_\lambda\psi_\lambda(t)} = \partial_\lambda\hat{H}(t)\ket{\psi_\lambda(t)}+\hat{H}(t)\ket{\partial_\lambda\psi_\lambda(t)},
$$
with $\ket{\partial_\lambda\psi_\lambda(0)} =0$. 

A generic measurement on the system is defined as a POVM of elements $\{\hat{E}_k\}$, with $\sum_k \hat{E}_k=\mathbb{I}$. Performing this measurement on the state $\ket{\psi_\lambda(t)}$ yields the result $k$ with probability $p_k(t)=\bra{\psi_\lambda(t)} \hat{E}_k\ket{\psi_\lambda(t)}$. The CFI associated with this measurement is defined as
$$
\mathcal{F}^{(C)}_\lambda=\sum_k\frac{1}{p_k(t)}\left(\frac{{\rm d} p_k(t)}{{\rm d} \lambda}\right)^2.
$$
We now define the subspace $\mathcal{H}_2$ generated by the vectors $\{\ket{\psi_\lambda(t)},\ket{\partial_\lambda\psi_\lambda(t)}\}$, which is of dimension 2 unless $\ket{\partial_\lambda\psi_\lambda(t)}\propto\ket{\psi_\lambda(t)}$, a case we exclude since it implies that the QFI is zero. From norm conservation, $\braket{\partial_\lambda\psi_\lambda(t)}{\psi_{\lambda}(t)}$ is imaginary, and we define in $\mathcal{H}_2$ the unitary vector $\ket{\psi_{\lambda,\perp}(t)}$ such that $\braket{\psi_{\lambda,\perp}(t)}{\psi_{\lambda}(t)}=0$, and $\braket{\psi_{\lambda,\perp}(t)}{\partial_\lambda\psi_\lambda(t)}$ is real. Hence we can write:
$$
\ket{\partial_\lambda\psi_\lambda(t)}=ib\ket{\psi_\lambda(t)}+c\ket{\psi_{\lambda,\perp}(t)},
$$
with $b,c\in \mathbb{R}$ and the QFI simply writes $\mathcal{F}_\lambda^{(Q)}=4c^2$.

We now consider the restrictions $\hat{E}_k^{(2)}$ of the measurement operators $\hat{E}_k$ to $\mathcal{H}_2$, which can generically be written in the orthonormal basis $\{\ket{\psi_\lambda(t)},\ket{\psi_{\lambda,\perp}(t)}\}$ in matrix form as:
$$
\left[\hat{E}_k\right]=\begin{bmatrix}
p_k & \beta_k \\
\beta_k^* & p_{k,\perp} 
\end{bmatrix},
$$
with $p_{k,\perp}=\bra{\psi_{\lambda,\perp}(t)} \hat{E}_k\ket{\psi_{\lambda,\perp}(t)}$ and $\beta_k$ a complex number. Since $\hat{E}_k^{(2)}$ remains a positive operator, we have $|\beta_k|\leq\sqrt{p_k,p_{k,\perp}}$. We denote $\phi_k=\arg{\beta_k}$. With these notations, the CFI takes the form:
$$
\mathcal{F}^{(C)}_\lambda=4c^2\sum_k \frac{|\beta_k|^2}{p_k}\cos^2(\phi_k).
$$

We deduce that \emph{coincidence $\mathcal{F}^{(C)}_\lambda=\mathcal{F}^{(C)}_\lambda$ is possible if and only if $\beta_k\in\mathbb{R}$ and $|\beta_k|=\sqrt{p_k,p_{k,\perp}}$ for all $k$}. 

\medskip

In the main text, we establish that the optimal control from the Fp strategy implements an evolution of $\ket{\psi_\lambda(t)}$ that follows a geodesic between the chosen Fp states $\ket{\psi_{1,2}}$. We can write this evolution generically as a function of $\lambda\in[\lambda_1,\lambda_2]$ :
$$
\ket{\psi_\lambda}=\cos{\frac{\theta(\lambda)}{2}}\ket{\psi_1}+\sin{\frac{\theta(\lambda)}{2}}\ket{\psi_2},
$$
where the argument $t$ was omitted for clarity. We then have $\ket{\partial_\lambda\psi_\lambda}=\tfrac{{\rm d}\theta}{{\rm d} \lambda}\ket{\psi_{\lambda_\perp}}$ along the trajectory. We now choose the measurement defined by $\hat{E}_1=\ketbra{\psi_1}{\psi_1}$, $\hat{E}_2=\ketbra{\psi_2}{\psi_2}$ and $\hat{E}_3=\mathbb{I}-\hat{E}_1-\hat{E}_2$. Simple calculations then show that the conditions for coincidence derived above are satisfied.

\section{QFI boundedness and $t^6$ scaling}
\label{scalingQFI}

Our proof in the main text that the optimal curve $\ket{\psi_\lambda(t_{\rm f})}$ is a geodesic relies on the QFI being bounded, which is not rigorously the case for the lattice system. 
However, the chosen control procedure with a fixed lattice depth can only transfer finite amounts of energy to the atoms. For a finite control time $t_{\rm f}$, the system is thus effectively confined in a finite-dimensional subspace, which allows us to draw the same conclusions.

We now present a heuristic proof of the time scaling of the QFI for force sensing with a BEC in a lattice. We first consider the Hamiltonian $\hat{H}(t)$ defined in Eq.~\eqref{eqSchro} as
$$
\hat{H}(t)=(\hat{p}-\lambda t)^2-\frac{s}{2}\cos(\hat{x}+\varphi(t)),
$$
and the subspace of $q=0$ generated by the states $\ket{\ell}$. Estimating the evolution of the QFI is complicated by the infinite dimension of the Hilbert space, denoted here by $\mathcal{H}$. In view of the previous point, we therefore introduce the finite dimensional Hilbert space $\mathcal{H}^{(N)}$ generated by the $2N+1$ eigenvectors of $\hat{p}$ such that $-N\leq \ell \leq N$ and the projector $P^{(N)}$ on this subspace, as well as the restriction of the Hamiltonian to $\mathcal{H}^{(N)}$, $\hat{H}^{(N)}=P^{(N)}\hat{H}P^{(N)}$.

Another equivalent formulation of QFI consists in introducing the operator $\hat{h}_\lambda(t)=\int_0^t\hat{U}(\tau,0)^\dagger \partial_\lambda\hat{H}(\tau) \hat{U}(\tau,0) d\tau$, where $\hat{U}(\tau,0)$ is the propagator from 0 to $\tau$ of the Schr\"odinger equation \eqref{eqSchro}. It can then be shown that~\cite{Pang_2017}
$$
\mathcal{F}^{(Q)}(t)=4\textrm{Var}[\hat{h}_\lambda(t)]|_{|\psi_0\rangle},
$$
where $\textrm{Var}[\cdot]$ is the variance of an operator. 

For the Hamiltonian considered here, we have $\partial_\lambda\hat{H}(\tau)=2\tau(\lambda \tau-\hat{p})$.  In the finite-dimensional Hilbert space $\mathcal{H}^{(N)}$, we obtain that $\mathcal{F}^{(Q)}\leq |\int_0^t(E_{\textrm{max}}^{(N)}-E_{\textrm{min}}^{(N)})d\tau|^2$ where $E_{\textrm{max}}^{(N)}$ and $E_{\textrm{min}}^{(N)}$ are respectively the maximum and minimum eigenvalues of $\partial_\lambda\hat{H}^{(N)}$ at time $\tau$, i.e. here $2\tau(\lambda\tau\pm N)$. We deduce that the QFI scales in $t^4$ or more precisely that $\mathcal{F}^{(Q)}\leq 4N^2 t^4$~\cite{Pang_2017}. The upper QFI bound is reached when $\ket{\psi_\lambda(\tau)}$ is equal at any time $\tau$ to a superposition with equal weights of the two corresponding eigenvectors $\ket{N}$ and $\ket{-N}$, $\ket{\psi_\lambda^{(N)}}=\tfrac{1}{\sqrt{2}}[\ket{N}+e^{i\phi(t)}\ket{-N}]$ with arbitrary phase $\phi(t)$. This bound cannot in general be saturated exactly by the dynamics, either because $\ket{\psi_0}$ cannot be expressed as a superposition of $\ket{\pm N}$, or because the system cannot be perfectly stabilized in the optimal state.

The $t^4$- scaling of the QFI is due to the finite dimension of the Hilbert space, but this limit can be overcome in an infinite-dimensional space. Based on the results of numerical optimization, we consider a specific control mechanism associated with a ladder-climbing process, in which the system is brought step by step from $\ket{\psi_\lambda^{(N)}}$ to $\ket{\psi_\lambda^{(N+1)}}$ in a fixed time $\delta$ with $N\delta=t$. We assume that $\delta$ is independent on $N$, as expected from the constant depth $s$ which provides a constant coupling between successive states. If the state of the system belongs to the subspace $\mathcal{H}^{(k+1)}$ in the time interval $[k\delta,(k+1)\delta]$ then we get
$$
\mathcal{F}^{(Q)}\leq \left[\sum_{k=0}^{N-1} \int_{k\delta}^{(k+1)\delta}
(E_\textrm{max}^{(k+1)}(\tau)-E_\textrm{min}^{(k+1)}(\tau)) d\tau \right]^2,
$$
with $N\delta = t$. Using $E_\textrm{max}^{(k+1)}(\tau)-E_\textrm{min}^{(k+1)}(\tau)=4\tau(k+1)$, we arrive at
$$
\mathcal{F}^{(Q)}\leq \left[\sum_{k=0}^{N-1}2\delta^2(k+1)(2k+1)\right]^2.
$$
For $N$ large enough, we have $\sum_{k=0}^N k^n\simeq N^{n+1}$. We deduce that
$$
\mathcal{F}^{(Q)}\leq \frac{16}{9}N^6\delta^4,
$$
where the lower orders in $N$ on the right hand side have been discarded. Finally, from $N\delta=t$, we obtain $\mathcal{F}^{(Q)}\leq \tfrac{16}{9}\tfrac{t^6}{\delta^2}$. We conclude that the time-scaling of the QFI is $t^6$, a rate that is much faster than the speed in a finite-dimensional Hilbert space~\cite{Pang_2017}.

The emerging single-fold and two-fold interferometer structures that are obtained from the optimal control (see Figs.~\ref{fig:theorie_Tstar} and \ref{fig:acc}) qualitatively behave as a succession of ladder ascents and descents, which will therefore yield the same scaling, albeit with a different prefactor.

\end{document}